\documentclass[12pt]{iopart}

\usepackage{iopams}
\usepackage{graphics,graphicx} 
\usepackage{multirow,rotate,color,float,times,dcolumn,bm,float}
\definecolor{blue}{rgb}{0,0,1}
\definecolor{red}{rgb}{1,0,0}

\begin{document}

\title[Social networks: models and measures]
      {New approaches to model and study social networks}

\author{P.G.~Lind}

\address{ICP, Universit\"at Stuttgart,
         Pfaffenwaldring 27, D-70569 Stuttgart, Germany}
\address{CFTC, Universidade de Lisboa, 
         Av.~Prof.~Gama Pinto 2, 1649-003 Lisbon, Portugal}

\author{H.J.~Herrmann}

\address{ICP, Universit\"at Stuttgart,
         Pfaffenwaldring 27, D-70569 Stuttgart, Germany}
\address{IfB, HIF E12, ETH H\"onggerberg, CH-8093 Z\"urich,
             Switzerland}
\address{Departamento de F\'{\i}sica, Universidade Federal do Cear\'a,
             60451-970 Fortaleza, Brazil}

\begin{abstract}
We describe and develop three recent novelties in network research 
which are particularly useful for studying social systems.
The first one concerns the discovery of some basic dynamical laws
that enable the emergence of the fundamental features observed
in social networks, namely the nontrivial clustering properties,
the existence of positive degree correlations and the subdivision
into communities.
To reproduce all these features we describe a simple model of
mobile colliding agents, whose collisions define the connections
between the agents which are the nodes in the underlying 
network, and develop some analytical considerations.
The second point addresses the particular feature of clustering
and its relationship with global network measures, namely with 
the distribution of the size of cycles in the network.
Since in social bipartite networks it is not possible to measure the 
clustering from standard procedures, we propose an alternative
clustering coefficient that can be used to extract an improved
normalized cycle distribution in any network.
Finally, the third point addresses dynamical processes occurring
on networks, namely when studying the propagation of information
in them.
In particular, we focus on the particular features of gossip propagation 
which impose some restrictions in the propagation rules.
To this end we introduce a quantity, the spread factor, which 
measures the average maximal fraction of nearest neighbors which 
get in contact with the gossip, and find the striking result that
there is an optimal non-trivial number of friends for which the spread 
factor is minimized, decreasing the danger of being gossiped.
\end{abstract}



\maketitle

\section{Introduction}

Contrary to what may be perceived at a first glance, social and physical
models were brought together several times, during the last four centuries.
In fact, not only Maxwell and Boltzmann were inspired by the statistical
approaches in social sciences to develop the kinetic theory of gases, but
one can even cite the English philosopher Thomas Hobbes, who already in the
seventeenth century, using a mechanical approach, tried to explain how 
people acquaintances and behaviors may contribute to the evolution towards 
a stable absolute monarchy\cite{ball02,ball03}.
More than making a historical perspective if these approaches were successful 
and correct or not, it is almost unquestionable that, at a certain level, 
there are social phenomena that could be more deeply understood by using 
approaches of statistical and physical models.
Recently\cite{barabasi02,newman03,dorogovtsev02}, 
this perspective gained considerable strength from the increased interest on - 
and in several senses well-succeed - network approach, where one describes 
complex systems by mapping them on a graph (network) of  nodes and links 
and studies their structure and dynamics with the help of some statistical 
and topological tools from statistical physics and graph 
theory\cite{boccaletti06,graphsbook}.

When addressing the specific case of a social system, nodes represent
individuals and the connections between them represent social relations
and acquaintances of a certain kind. 
Social networks were studied in different contexts%
\cite{newman01,newman02,krapivsky03,rogers03,freemanbook,prl,epjb}, 
ranging from epidemics spreading and sexual contacts to language evolution 
and vote elections.
However, although they are ubiquitous, social networks differ from most
other networks, yielding a still broad spectrum of unanswered questions and
improvements to be done when studying their statistical and topological 
properties. 
In this paper, we will address three fundamental open questions related
to the typical structure and dynamics associated to social networks.

The first open question has to do with the modeling of social networks.
The recent broad study of empirical social networks has shown that they
have three fundamental features common to all of them\cite{newman03b}. 
First, they present the
small-world effect\cite{watts98} with small average path lengths between 
nodes and high clustering coefficients meaning that neighbors tend to be 
connected with each other. Second, they have positive correlations:
the highly (poorly) connected nodes tend to connected to other highly
(poorly) connected nodes. Third and last, invariably one observes an
organization of the network into some subsets of nodes (communities) more
densely connected between each other.
Although there are arguments pointing out that all these features could be
consequence from one another\cite{newman03b}, 
the modeling of specific social networks
reproducing quantitatively all these features has not been successful.
Using a recent approach to construct networks, based on a system of mobile
agents, it is possible to reproduce all these features.
In Section \ref{sec:mobileagents} we will further show that the 
degree distributions characterizing social networks typically follow
a specific one-parameter distribution, so-called Brody distribution.

The second question is related to the intrinsic nature of the nodes.
For certain social networks there are intrinsic features of the individuals
which must be considered in the analysis. For instance, the gender in
networks of sexual contacts\cite{epjb} or the hierarchical position
in a network of social contacts inside some enterprise.
From the network point of view this distinction means to introduce 
multipartivity in the network, biasing the preferential attachment
between nodes that tend to connect with nodes of a certain type.
When there are two types of nodes, e.g.~men and women, and the connections
between them is strongly related to this type, e.g.~men can only match 
women and vice-versa, the standard measures
to analyze network structure fails. In particular, the standard clustering
coefficient\cite{watts98}, is unable to quantify the 
connectedness of broader neighborhoods that typically appear in multipartite 
networks.
In Sec.~\ref{sec:clustering} we will revisit some of the clustering 
coefficients used to study clustering in bipartite networks, and show
how the combination of both clustering coefficients can yield good
estimates of normalized cycle distributions.
Moreover, we will discuss a general theoretical picture of a global measure
of increasing order of clustering coefficients according to some
suitable expansion.

The third open question has to do with the heterogeneity of nodes in
what concerns their influence in the connections  and therefore in the 
propagation phenomena on social networks. 
In rumor propagation\cite{daley64}, for instance, one usually treats all 
connections
equally in the spread of some signal (opinion, rumor, etc). 
This is a suitable assumption for situations like the spread of an opinion
which is equally interesting to all nodes in the network, for example
political opinions to some vote election.
However, there are also several social situations where the signal is
not equally interesting to all nodes, as the case of spreading of some
gossip about some common friend.
In these cases there are connections which will be more probably used to 
spread the signal than others, since not all our friends are also friends
of the particular person which is being gossiped about and therefore, 
either we tend to not tell the gossip to them or they tend to not spread 
it even if they hear it.
In Sec.~\ref{sec:gossip} we will present a simple model for gossip 
propagation and describe some striking features.
Namely, that there is an optimal number of friends, depending on the degree 
distribution and degree correlations of the entire network, for which the 
danger to be gossiped is minimized.

Finally, in Sec.~\ref{sec:conclusions} we make final conclusions,
giving an overview of future questions which could be studied 
in social networks arising from the topics studied throughout the
paper.

\section{Modeling social networks: 
         an approach based on mobile agents}
\label{sec:mobileagents}

Since the study of social networks is mainly concerned with
topological and statistical features of people's 
acquaintances\cite{newman01,newman02},
the modeling of such networks has been done within the
framework of graph theory using suitable probabilistic
laws for the distribution of connections between 
individuals\cite{barabasi02,newman03,dorogovtsev02,graphsbook}.
This approach proved to be successful in several contexts,
for instance to describe community formation\cite{groenlund04,toivonen06}
and their growth\cite{jin01}.
\begin{figure}[t]
\begin{center}
\includegraphics*[width=10.0cm]{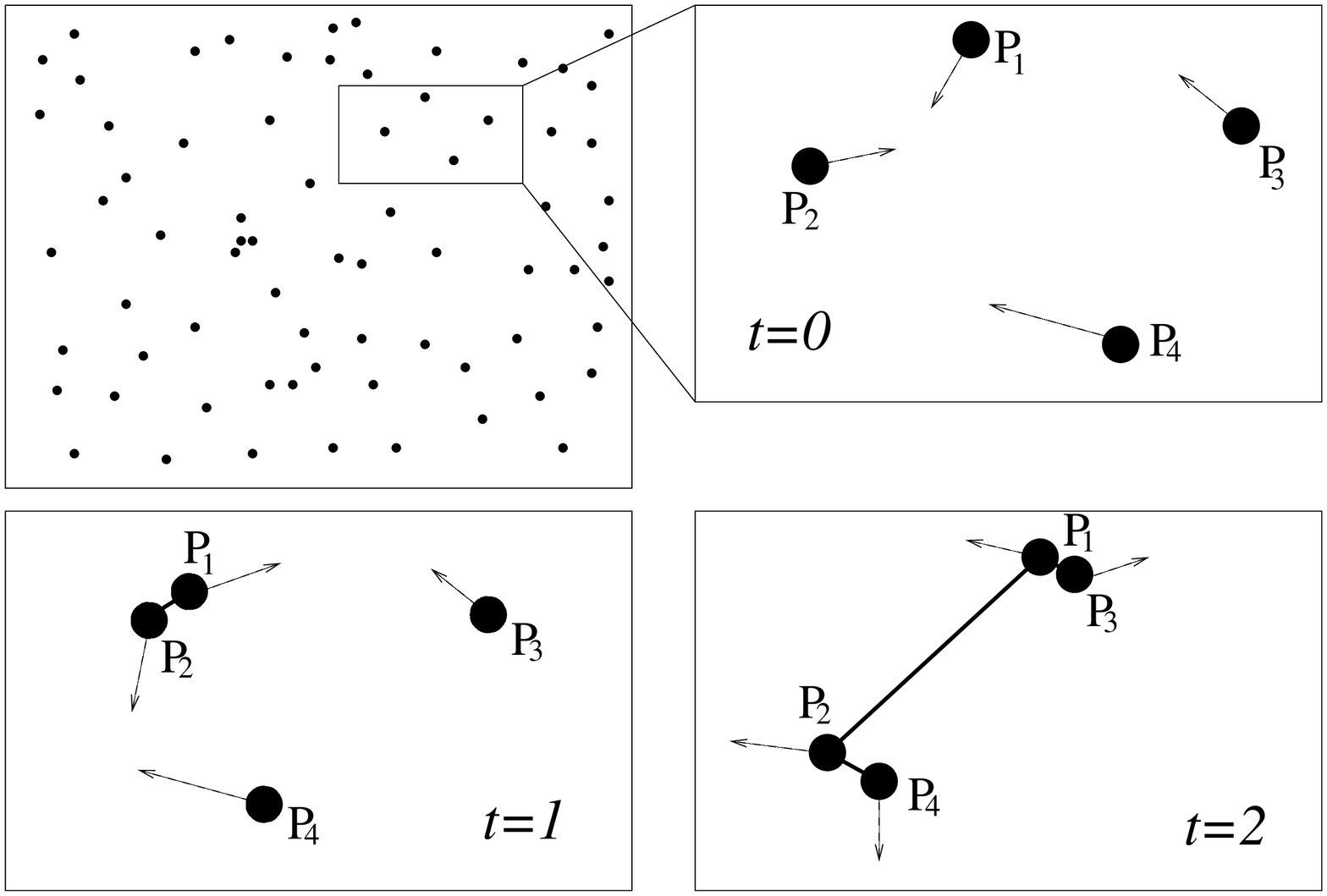}
\end{center}
\caption{\protect
         Illustration of the two-dimensional mobile agents system. 
         Initially there are no connections between nodes and nodes 
         move with some initial velocity $v_0$ in a randomly chosen
         direction (arrows).
         At $t=1$ two nodes, $P_1$ and $P_2$ collide and a connection 
         between them is introduced (solid line), velocities are updated
         increasing their magnitude and choosing a new random direction.
         At $t=2$  two other collisions occur, between nodes $P_2$ and $P_4$
         and between nodes $P_1$ and $P_3$.
         In this way a network of nodes and connections between them
         emerges as a straightforward consequence of their motion 
         (see text).}
\label{fig01}
\end{figure}

However, they present two major drawbacks. First, the graph approach
may be suited to describe the structure of social contacts and
acquaintances, but lacks to give insight into the social dynamical laws
underlying it. Second, these models seem to be unable to reproduce
all the main features characteristic of social networks, at least at
the fundamental level.
In this context, it was pointed out that\cite{pre,davidsen02,eisenberg03}
dynamical processes based on local information should be also considered
when modeling the network.
Our recent proposal to overcome these shortcomings was to construct
networks, from a system of mobile agents following a simple motion
law\cite{prl,epjb}. Here, we briefly review this model and further present 
the analytical expression that fits the obtained degree distributions.
In particular we show that the degree distribution typically follows a 
Brody distribution\cite{brody73}.

\subsection{The model}
\label{subsec:agmodel}

The model is given by a system of particles (agents) that move
and collide with each other, forming through those collisions
the acquaintances between individuals.
Consequently, the network results directly from the time evolution of
the system and is parameterized by two single 
parameters, the density $\rho$ of agents characterizing the system
composition and the maximal residence time $T_{\ell}$ controlling its 
evolution.
Each agent $i$ is characterized by its number $k_i$ of links and by its age 
$A_i$. 
When initialized each agent has a randomly chosen age, position and moving 
direction with velocity $v_0$ and one sets $k_i=0$.
While moving, the individuals follow ballistic trajectories till they
collide.
As a first approximation we assume that social contacts do not determine 
which social contact will occur next. 
Therefore, after collisions, the total momentum should not be conserved, 
with the two agents choosing completely random new moving directions.
Figure \ref{fig01} sketches consecutive stages of the evolution of
such a system of mobile agents.

Assuming that large number of acquaintances tend to favor the occurrence
of new contacts, the velocity should increase with degree $k$, namely
\begin{equation}
\vec{v}(k_i)=(\bar{v}k_i^{\alpha}+v_0)\vec{\omega} ,
\label{velo}
\end{equation}
where $\bar{v}=1\hbox{\ m/s}$ is a constant to assure dimensions of 
velocity, 
$\vec{\omega}=(\vec{e}_x\cos{\theta}+\vec{e}_y\sin{\theta})$ with 
$\theta$ a random angle and $\vec{e_x}$ and $\vec{e_y}$ are unit vectors.
The exponent $\alpha$ in Eq.~(\ref{velo}) controls the velocity update  
after each collision.
Here, we consider $\alpha=1$.
Further, the removal of agents considered here is simply imposed by some
threshold $T_{\ell}$ in the age of the agents: when $A_i=T_{\ell}$
agent $i$ leaves the system and a new agent $j$ replaces it with
$k_j=0$, $v_j=v_0$ and randomly chosen moving direction.
The selected values for $T_{\ell}$ must be of the order of several
times the characteristic time $\tau$ between collisions, in order to 
avoid either premature death of nodes.
Too large values of $T_{\ell}$ are also inappropriate since in that case
each node may on average collide with all other nodes yielding a fully
connected network.
\begin{figure}[t]
\begin{center}
\includegraphics*[width=10.0cm]{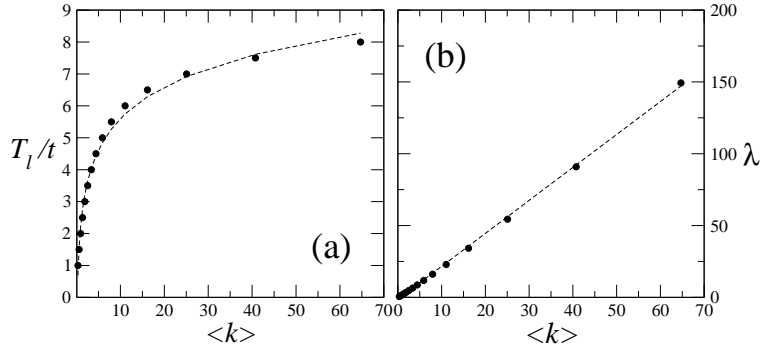}
\end{center}
\caption{\protect
        Bridging between real social networks with average
        degree $\langle k\rangle$ and the system of mobile 
        agents that reproduce their topological and statistical
        features.
        In {\bf (a)} the normalized maximal residence time of
        agents is plotted as a function of the average degree,
        while in
        {\bf (b)} one plots the collision rate $\lambda$ which
        is a unique function of the residence time, and
        scales with $\langle k\rangle$.} 
\label{fig02}
\end{figure}

Similarly to other systems\cite{Amaral,dorogovtsev00},
this finite $T_{\ell}$ enables the entire system
to reach a non-trivial quasi-stationary state\cite{prl}.
In fact, only by tuning $T_{\ell}$ within an
acceptable range of small density values, one reproduces networks
of social contacts.
In Fig.~\ref{fig02}a one sees the normalized residence time
$T_{\ell}/\tau$ as a strictly monotonic function of the average degree 
$\langle k\rangle$.
From the residence time it is also possible to define a collision 
rate, as the fraction between the average residence time 
$T_{\ell}-\langle A(0)\rangle=T_{\ell}/2$ and the
characteristic time $\tau$, namely
$\lambda=T_{\ell}/(2\tau)=\langle v\rangle T_{\ell}/(2v_0\tau_0)$,
where $\tau_0$ is the characteristic time of the system at the beginning
when all agents have velocity $v_0$.
Figure \ref{fig02}b shows clearly that $\lambda=2\langle k\rangle$.

By looking at Fig.~\ref{fig02} one now understands the main strength
of the mobile agent model here described: when taking a real network
of social contacts and measuring the average degree $\langle k\rangle$
the correspondence sketched in Fig.~\ref{fig02} straightforwardly
returns the suitable value of $T_{\ell}$ that reproduces the topological and
statistical features.
\begin{figure}[t]
\begin{center}
\includegraphics*[width=13.0cm]{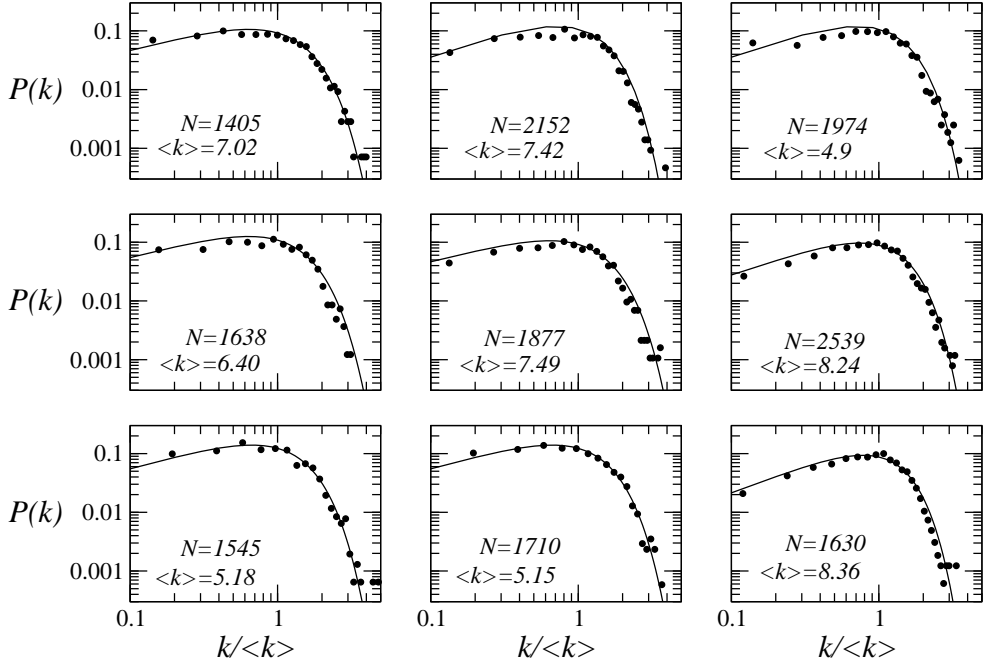}
\end{center}
\caption{\protect
        Degree distributions of nine different schools (symbols) from an
        in-school questionnaire involving a total of $90118$ students which 
        responded to it in a survey between 1994 and 1995. 
        Each school comprehends a number $N$ of interviewed students and
        from their questionnaires an average number $\langle k\rangle$ of 
        acquaintances is extracted.
        With solid lines we represent the fit obtained with a Brody
        distribution, Eq.~(\ref{brody}), whose parameter value is computed
        in Fig.~\ref{fig04}.}
\label{fig03}
\end{figure}

It was already reported\cite{prl,physicaD} that empirical networks
extracted from a survey among $84$ American schools are easily reproduced
with this mobile agent model, in what concerns the degree distribution,
second-order correlations, community structure, average path length
and clustering coefficient.
As an illustration, Fig.~\ref{fig03} shows the degree distribution $P(k)$ 
of nine of such schools (symbols).
Such distributions are well fitted by Brody distributions
(solid lines) defined as\cite{brody73}:
\begin{equation}
P_B(\bar{k}) = \frac{1}{B}(\beta+1)\eta \bar{k}^{\beta} 
\exp{(-\eta \bar{k}^{\beta+1})},
\label{brody}
\end{equation}
with $\bar{k}=k/\langle k\rangle$ and
\begin{equation}
\eta = \Gamma \left ( \frac{\beta+2}{\beta+1}\right )^{\beta+1}.
\end{equation}
and $B$ a normalization constant.
Roughly, the Brody distribution in Eq.~(\ref{brody}) is, apart some
special constants, the product of a power of $k$ with an exponential 
with a negative exponent proportional to a higher power of $k$.
For the particular case $\beta=0$, the Brody distribution reduces to
the exponential distribution having always a non-positive derivative. 

\begin{figure}[t]
\begin{center}
\includegraphics*[width=6.0cm]{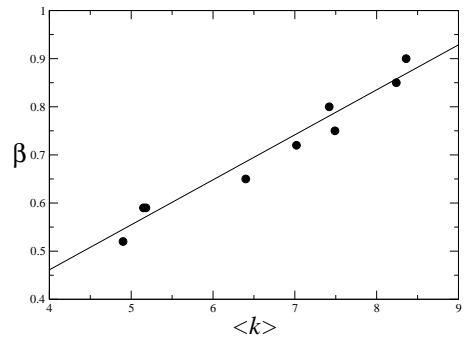}
\end{center}
\caption{\protect
        The linear dependence between the parameter $\beta$ of the
        Brody distribution in Eq.~(\ref{brody}) with the average
        number $\langle k\rangle$ of connections. Each bullet corresponds
        to one of the schools whose degree distribution is plotted 
        in Fig.~\ref{fig03}. The solid line yields the fit
        $\beta=0.094\langle k\rangle + 0.078$.}
\label{fig04}
\end{figure}

The distributions in Fig.~\ref{fig03} were obtained with
values of $\beta$ slightly above zero, namely between zero and one
as shown in Fig.~\ref{fig04}.
In this case one is able to obtain the non trivial positive slope
which is typically observed for small $k$ values in the degree distribution
of such social networks.
Interestingly, Fig.~\ref{fig04} also shows a linear trend between the average
degree $\langle k\rangle$ in the network and the corresponding value of $\beta$ 
which fits the degree distribution. This guarantees that distribution in 
Eq.~(\ref{brody}) has indeed one single parameter.
How such a distribution can be obtained from an analytical approach
to the model of mobile agents is still an open question and will be
discussed in detail elsewhere.

\section{Particular measures for social networks}
\label{sec:clustering}

To measure ``the cliquishness of a typical neighborhood'' in a network,
Watts and Strogatz~\cite{watts98} introduced a simple coefficient,
called the clustering coefficient,
which counts the number of pairs of neighbors of a certain node
which are connected with each other, forming a cycle of size $s=3$.
While such tool enables one to access the structure of complex networks 
arising in many systems~\cite{newman03,graphsbook}, helping to 
characterize small-world networks~\cite{watts98}, 
to understand synchronization in scale-free
networks of oscillators~\cite{mcgraw05} and to characterize chemical
reactions~\cite{gleiss01} and networks of social
relationships~\cite{newman03a,holme1}, there are other situations
where this measure does not suit.
Namely, when the network presents a multipartite structure.
For instance, when there are two different kinds of nodes 
and connections link only nodes of different type, the network
is bipartite\cite{newman03a,holme1,holme3} and the bipartite structure 
does not allow the occurrence of cycles with odd size, in particular 
with $s=3$.

Bipartite networks are quite common for social
systems\cite{holme3,guimera} where the two different kinds
of nodes represent e.g.~the two genders.
While the standard clustering coefficient in such networks is always
zero, they have in general non vanishing clustering properties\cite{holme1} 
and therefore more appropriate quantities to access such networks have 
been proposed, namely coefficients counting larger cycles.
In this Section, we will discuss how these different clustering coefficients
are related to each other and how one can use them to improve the knowledge
of the network structure.

The standard clustering coefficient $C_3$ is usually defined\cite{watts98} 
as the fraction between the number of cycles of size $s=3$ (triangles) 
observed in the network out of the total number of possible triangles 
which may appear, namely
\begin{equation}
C_3(i) = \frac{2t_i}{k_i(k_i-1)} . \label{c3}
\end{equation}
where $t_i$ is the number of existing triangles containing node $i$
and $k_i$ is the number of neighbors of node $i$, yielding a maximal
number $k_i(k_i-1)/2$ of triangles.

To access the cliquishness in bipartite networks one
has proposed\cite{pre,holme1,holme3,gridref} a clustering coefficient 
$C_4(i)$, sometimes called the grid coefficient\cite{gridref}, 
defined as the quotient between the number of cycles of size $s=4$ 
(squares) and the total number of possible squares.
Explicitly, for a given node $i$ with two neighbors, say $m$ and $n$, 
this coefficient yields\cite{pre}
\begin{equation}
C_{4,mn}(i) = \frac{q_{imn}}
               {(k_m-\eta_{imn})
                (k_n-\eta_{imn})+q_{imn}},
\label{c4}
\end{equation}
where $q_{imn}$ is the number of common neighbors between $m$ and $n$
(not counting $i$) and $\eta_{imn}=1+q_{imn}+\theta_{mn}$ with
$\theta_{mn}=1$ if neighbors $m$ and $n$ are connected with each other 
and $0$ otherwise. 

After averaging over the nodes, the coefficients $C_3$ and $C_4$
characterize the contribution of the first and second neighbors,
respectively, for the network cliquishness.
In order to be a suitable quantity to measure the 
cliquishness of bipartite networks compared to their monopartite 
counterparts, $C_4$ must behave the same way as $C_3$ when the network 
parameters are changed, as it is indeed the case for $\langle C_4\rangle$
computed from Eq.~(\ref{c4}). See Ref.~\cite{pre} for details.

One should notice that in most $m$-partite networks, 
it is always possible to have cycles of size $s=4$, indicating
that $C_4$ is in some sense a more general clustering measure than $C_3$.
However, it could be the case that for a larger number of partitions 
forming the network, the contribution of larger cycles increases.
This is the case, for instance, of trophic relations in an ecological
network of different individuals from different species, where large
cycles tend to be abundant, namely the ones ranging from the higher
predators to the plants at the lowest trophic level.  
In such cases, a general clustering coefficient counting the
fraction of possible cycles of arbitrary size $n$ may be needed.
The generalization is straightforward yielding a clustering
coefficient $C_n = E_n/L_n$,
where $E_n$ is the number of existing cycles with size $n$, $L_n$
the maximal number of such cycles that is possible to be attained and
$n=3,\dots,N$ for a network of $N$ nodes.

Having $C_n$ for the required values of $n$, one is able to introduce a 
general clustering measure of the network, given by the sum of all these 
contributions, namely
\begin{equation}
C = \sum_{n=3}^N \alpha_n C_n = \sum_{n=3}^N \alpha_n \frac{E_n}{L_n} ,
\label{ccexpans}
\end{equation}
where $\alpha_n$ is a coefficient that weights the contribution of each
different clustering order $n$ and obeys the normalization condition
$\sum_{n=3}^N \alpha_n = 1$.
In general one can write $E_n$ and $L_n$ in Eq.~(\ref{ccexpans}) as
\begin{eqnarray}
E_n &=& \sum_{k_1,\dots ,k_n} NP(k_1)q(k_1,k_2)NP(k_2)q(k_2,k_3)\dots
                              NP(k_{n-1})q(k_{n-1},k_n) \label{en}\\
L_n &=& B^{N}_{n} n! = \frac{N!}{(N-n)!} \label{ln}
\end{eqnarray} 
where $B^N_n$ are the total combinations of $n$ elements out of $N$,
$P(k)$ is the fraction of nodes with $k$ neighbors and $q(k_1,k_2)$ is 
the correlation degree distribution, i.e.~the fraction of connections linking
a node with $k_1$ neighbors to a node with $k_2$ neighbors.

From Eq.~(\ref{en}) one can assume approximately that $E_n\sim 
(\langle P\rangle\langle q\rangle N)^n$ with $\langle P\rangle$ and
$\langle q\rangle$ the average fractions of $P(k)$ and $q(k_1,k_2)$
respectively.
Since $L_n$ increases also as $N^n$, a possible suitable choice
for $\alpha$ would be a constant, namely $\alpha = 1/(N-2)$ obeying
the normalization condition above.
Having presented this general scenario, we now concentrate on the two 
first clustering coefficients, $C_3$ and $C_4$, to address the cycle
size distribution.
\begin{figure}[t]
\begin{center}
\includegraphics*[width=8.5cm,angle=0]{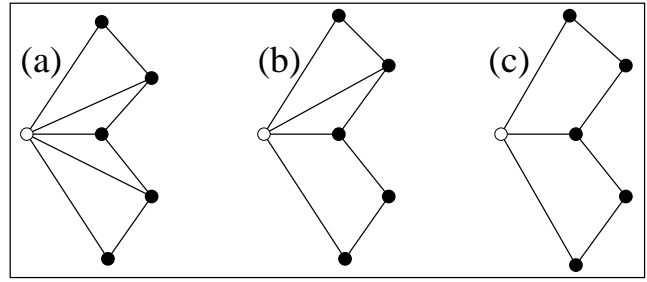}
\end{center}
\caption{\protect
         Illustrative examples of cycles (size $s=6$) where the most
         connected node ($\circ$) is connected to
         {\bf (a)} all the other nodes composing the cycle, forming
         four adjacent triangles.
         In {\bf (b)} the most connected node is connected to all
         other nodes except one, forming two triangles and one
         sub-cycle of size $s=4$, while in
         {\bf (c)} the same cycle $s=6$ encloses two sub-cycles of
         size $s=4$ and no triangles (see text).}
\label{fig05}
\end{figure}

We first show an estimate 
introduced in Ref.~\cite{barabasi05}, which 
considers only the degree distribution $P(k)$ and the distribution
of the standard  clustering coefficient $C_3(k)$.
One starts by considering the set of cycles with a central node, i.e.~cycles 
with one node connected to all other nodes composing the cycle, as
illustrated in Fig.~\ref{fig05}a.
The central node composes one triangle with each pair of connected neighbors.
Due to this fact, the number of cycles with size $s$ can be easily
estimated, since the number of different possible cycles to occur is 
$n_0(s,k)=B^k_{s-1}\frac{(s-1)!}{2}$, for a central node with $k$ neighbors
and the corresponding fraction of these cycles which is expected to occur is
$p_0(s,k)=C_3(k)^{s-2}$, yielding a total number of $s$-cycles given by
\begin{eqnarray}
N_s &=& Ng_s \sum_{k=s-1}^{k_{max}}
      P(k)n_0(s,k)p_0(s,k) , 
\label{estimate_c3}
\end{eqnarray}
where $g_s$ is a factor which takes into account the number of  
cycles counted more than once.

The estimate in Eq.~(\ref{estimate_c3}) is a lower bound for the
total number of cycles since it considers only cycles with a central
node.
Further, this estimate only accounts for cycles up to size
$s\le k_{max}+1$, with $k_{max}$ the maximal degree and 
is not suited for bipartite networks where
$C_3(k)=0$ for all $k$. Bipartite networks are typically
composed of a set of nodes as those illustrated in Fig.~\ref{fig05}c,
where no central node exists.

By using additionally the coefficient $C_4(k)$ in a similar estimate,
one is now able to take into account several cycles without central nodes.
One first considers the set of cycles of size $s$ with one node
connected to all the others {\it except} one, as illustrated
in Fig.~\ref{fig05}b. 
Assuming that this node has $k$ neighbors, $s-2$ of them belonging
to the cycle one is counting for, one has $n_1(s,k)=B^k_{s-2}(s-2)!/2$ 
different possible cycles of size $s$.
The corresponding fraction of such cycles which is expected to occur
is given by $p_1(s,k)=C_3(k)^{s-4}C_4(k)(1-C_3(k))$.
Writing an equation similar to Eq.~(\ref{estimate_c3}), where instead 
of $n_0(s,k)$ and $p_0(s,k)$ one has $n_1(s,k)$ and $p_1(s,k)$ respectively
and the sum starts at $s-2$ instead of $s-1$, one has an additional number
$N^{\prime}_s$ of estimated cycles which is not considered in estimate
(\ref{estimate_c3}).

To improve the estimate further one repeats the same  approach, taking out
each time one connection to the initial central node, increasing 
by one the number of elementary cycles of size $s=4$.
Figure \ref{fig05}c illustrates a cycle of size $s=6$ composed by two
elementary cycles of size $4$.
In general, for cycles composed by $q$ sub-cycles of size $4$ one finds
$n_q(s,k)=\frac{(s-q-1)!}{2} B^k_{s-q-1}$ possible cycles of size
$s$ looking from a node with $k$ neighbors and 
a fraction $p_q(s,k)=C_3(k)^{s-2q-2}C_4(k)^{q} (1-C_3(k))^{q}$ of them 
which are expected to be observed.

Summing up over $k$ and $q$ yields our final expression
\begin{equation}
N_s = Ng_s \sum_{q=0}^{[s/2]-1}\sum_{k=s-q-1}^{k_{max}}
      P(k)n_q(s,k)p_q(s,k) .
\label{estimate_tot}
\end{equation}
where $[x]$ denotes the integer part of $x$.
In particular, the first term ($q=0$) is the sum in Eq.~(\ref{estimate_c3})
and the upper limit $[s/2]-1$ of the first sum is obtained by
imposing the exponent of $C_3(k)$ in $p_q(s,k)$ to be non-negative.

The estimate in Eq.~(\ref{estimate_tot}) not only improves the estimated
number computed from Eq.~(\ref{estimate_c3}), but also enables the estimate 
of cycles up to a larger maximal size\cite{pre}, namely up to $s=2k_{max}$
where $k_{max}$ is the maximal number of neighbors in the network.
\begin{figure}[b]
\begin{center}
\includegraphics*[width=8.5cm,angle=0]{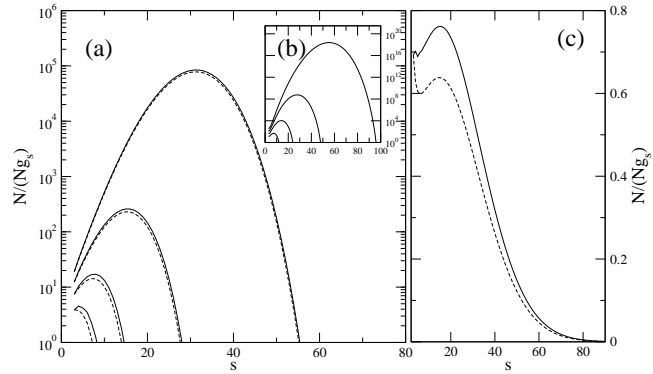}
\end{center}
\caption{\protect
    {\bf (a)} The fraction $N_s/Ng_s$ of the number of cycles
    estimated from Eqs.~(\ref{estimate_c3}), dashed lines, and 
    (\ref{estimate_tot}), solid lines, compared with 
    {\bf (b)} the exact number of cycles as a function of the size for 
    the pseudo-fractal network~\cite{dorog}.
    From small to large curves one has pseudo-fractal
    networks with $m=2,3,4,5$ generations (see text).
    In {\bf (c)} one sees the comparison between both estimates
    in a scale-free network with degree distribution $P(k)=P_0 k^{-\gamma}$
    with $P_0=0.737$ and $\gamma=2.5$, and coefficient
    distributions $C_{3,4}(k)=C_{3,4}^{(0)}k^{-\alpha}$ with
    $C_3^{(0)}=2$, $C_4^{(0)}=0.33$ and $\alpha=0.9$.}
\label{fig06}
\end{figure}

The estimate in Eq.~(\ref{estimate_tot}) has also the advantage of
being able to estimate cycles in bipartite networks.
Since for bipartite networks $C_3(k)=0$,
all terms in Eq.~(\ref{estimate_tot}) vanish except those
for which the exponent of $C_3(k)$ is zero, i.e.~for $s=2(q+1)$ with
$q$ an integer, which naturally shows the absence of cycles of 
odd size in such networks.

For highly connected networks, both estimates should nevertheless
yield similar results, since in that case there is a very large number 
of both triangles and squares.
For instance, the so-called pseudo-fractal network\cite{dorog}
is a deterministic scale-free network, constructed from three 
initial nodes connected with each other (generation $m=0$),
and iteratively adding new generations of nodes such that
in generation $m+1$ one new node is added to each edge of generation
$m$ and is connected to the two nodes joined by that edge.
For these networks, the exact number of cycles with size $s$ 
can be written iteratively~\cite{klemm05} and 
can be directly compared to the one obtained
with the two estimates above.
Figure \ref{fig06}a shows the two estimates, while in Fig.~\ref{fig06}b
the exact number is computed. 
We notice that both the real number $N_s$ of cycles and the normalized 
value $N_s/(Ng_s)$, though different, yield the same shape.
Thus, although the estimates above are not able to explicit the geometrical
factor $g_s$, the corresponding normalized distributions 
agree very well with the real one. 
However, while in this simple situation both estimates are similar, 
in general they can deviate significantly, as illustrated in 
Fig.~\ref{fig06}c. In such cases, the estimate (\ref{estimate_tot}) is 
closer to the real distribution of cycle sizes\cite{pre}.

\section{Spreading phenomena in social networks}
\label{sec:gossip}

In the previous Section we show how the study of network structure can be 
addressed by using tools as the clustering coefficient and first and second
degree distributions.
However, although the ability to communicate within a network of contacts
is favored by the network topology\cite{trusina05}, to study dynamical 
phenomena occurring on the network other measures are necessary.
Here, we focus on novel properties that help to ascertain
the broadness and speed of propagating phenomena through the network.
We will describe two helpful
quantities to study propagation in a network.
As we will see these tools are particularly suited for a simple
model of gossip propagation, that yields a striking result:
in real social systems it is possible to minimize the 
risk of being gossiped, by only choosing an optimal number of friendship 
acquaintances.

We start by introducing the additional quantities in the context of gossip
propagation.
As opposed to rumors, a gossip always targets the details about
the behavior or private life of a specific person.
Some information of a specific gossip is created at time
$t=0$ about the victim by one of its neighbors.
Since typically the gossip tends to be of interest to only those
who know the victim personally, we 
consider first that it only spreads at each time step from the 
vertices that know the gossip to all vertices that are connected 
to the victim and do not yet know the gossip. 
Our dynamics is therefore like a burning algorithm~\cite{burning}, 
starting at the originator and limited to sites that are neighbors 
of the victim. The gossip will spread until all reachable neighbors 
of the victim know it, yielding a spreading time $\tau$.

To measure how effectively the gossip or more generally the amount of 
information attains the neighbors of the starting node (victim), we 
define the spreading factor $f$ given by
\begin{equation}
f=n_f/k
\end{equation}
where $n_f$ is the total number of the $k$ neighbors who eventually 
hear the gossip in a network with $N$ vertices (individuals).
Although similar in particular cases, the spreading factor $f$ 
and the clustering coefficient are, in general, different because the 
later one only 
measures the number of bonds between neighbors giving no insight about how
they are connected.
\begin{figure}[t] 
\begin{center}
\includegraphics*[width=8.5cm,angle=0]{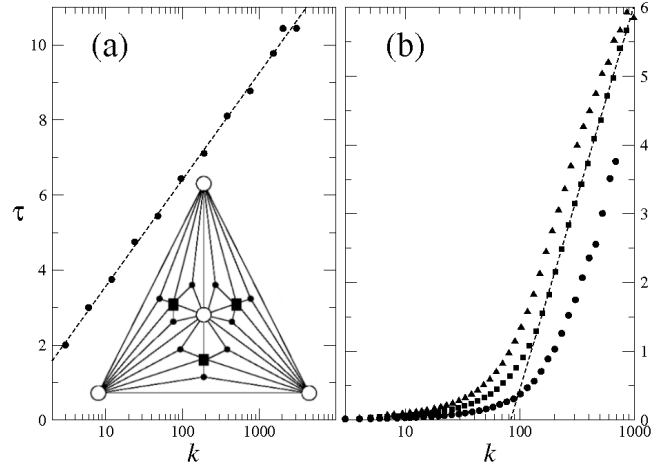}
\end{center}
\caption{\protect 
   Semi-logarithmic plot of the spreading time $\tau$ as a function of
   the degree $k$ for (a) the Apollonian ($n=9$ generations) and (b)
   the Barab\'asi-Albert network with $N=10^4$ nodes for $m=3$
   (circles), $5$ (squares) and $7$ (triangles), where $m$ is the
   number of edges of a new site, and averaged over $100$
   realizations. In the inset of (a) we show a schematic design of the 
   Apollonian lattice for $n=3$ generations. Fitting Eq.~(\ref{eq1}) 
   to these data we have $B = 1.1$ in (a) and $B = 5.6$ for large $k$ 
   in (b).}
\label{fig07}
\end{figure}

In Fig.~\ref{fig07} one sees how the spreading time $\tau$ depends on the
degree $k$ of the starting node. The Apollonian network\cite{hans} 
is illustrated in Fig.~\ref{fig07}a, while the case of Barab\'asi-Albert 
networks is given in Fig.~\ref{fig07}b.
In both cases $\tau$ clearly grows logarithmically, 
\begin{equation}
\tau = A + B {\rm log}k ,
\label{eq1}
\end{equation}
for large $k$. In the case of the Apollonian
network, one can even derive this behavior analytically as follows. 
In order to communicate between two vertices of the $n$-th generation, 
one needs up to $n$ steps, which leads to
$\tau \propto n$. Since for the Apollonian network one has\cite{hans} 
$k=3 \times 2^{n-1}$, one immediately obtains that $\tau \propto {\rm log}k$. 

For the Apollonian network all neighbors of a given victim 
are connected in a closed path surrounding the victim, 
as can be seen from the inset of Fig.~\ref{fig07}a, yielding $f=1$.
This stresses the fact that the spread factor $f$ is rather different 
from the clustering coefficient which in this case is 
$C=0.828$~\cite{hans}. 

\begin{figure}[t]
\begin{center}
\includegraphics*[width=8.5cm,angle=0]{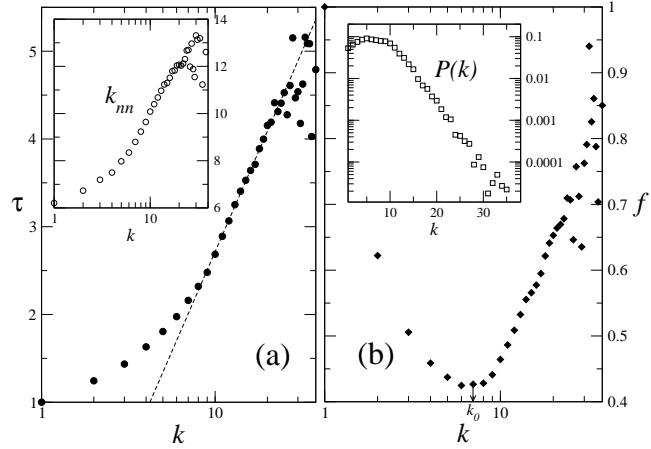}
\end{center}
\caption{\protect 
        Gossip propagation on a real friendship network of American  
        students~\cite{schools} averaged over 84 
        schools. In (a) we show the spreading time
        $\tau$ and, in the inset, the average degree of neighbors
        of nodes with degree $k$.
        In (b) the spread factor $f$, both as a function of
        degree $k$. In the inset of (b) we see the degree distribution
        $P(k)$.}
\label{fig08}
\end{figure}

Next, we will show that for these two features to appear one needs 
the existence of degree correlations between connected nodes, 
as usually observed in real empirical networks.
In Fig.~\ref{fig08} we plot the results of gossip spreading
on an empirical set of networks extracted from survey data\cite{schools}
in 84 U.S. schools.
Here, the logarithmic growth of $\tau$ with $k$, shown in 
Fig.~\ref{fig08}a, follows the same dependence of the average degree 
$k_{nn}$ of the nearest neighbors\cite{catanzaro}, as illustrated in 
the inset.
As in the case of the
BA networks, we also find for the schools a characteristic
degree $k_0$ for which $f$ and therefore the gossip spreading is
smallest. 
The inset of Fig.~\ref{fig08}b, however, gives 
clear evidence that the school networks are not scale-free.
Since the same optimal degree appears in Barab\'asi-Albert networks,
one argues that the existence of this optimal number
is not necessarily related to the degree distribution of the network,
but rather to the degree correlations.

However, the relation between degree correlations, measured
by $k_{nn}$, and the logarithmic behavior of the spreading time is
not straightforward. While in the empirical network we found the same
distribution for both $k_{nn}$ and $\tau$, in BA and APL networks
$k_{nn}$ follows a power-law with $k$ (not shown).
As for the spread factor $f$, a mean field approach can be derived,
yielding an $f$-rate equation which depends in general on $P(k)$ and
two and three-point correlations of the degree.
In the case of uncorrelated networks, two and three-point correlations
reduce to simple expressions of the moments of the degree distribution.
Therefore, $f$ is independent of the degree, similarly to what is
observed for the density of particles as derived by Catanzaro et 
al\cite{catanzaro2} in diffusion-annihilation processes on complex 
networks.
For correlated networks, as the empirical network here studied, the 
analytical approach is not straightforward and will be presented elsewhere.
\begin{figure}[t]
\begin{center}
\includegraphics*[width=8.5cm,angle=0]{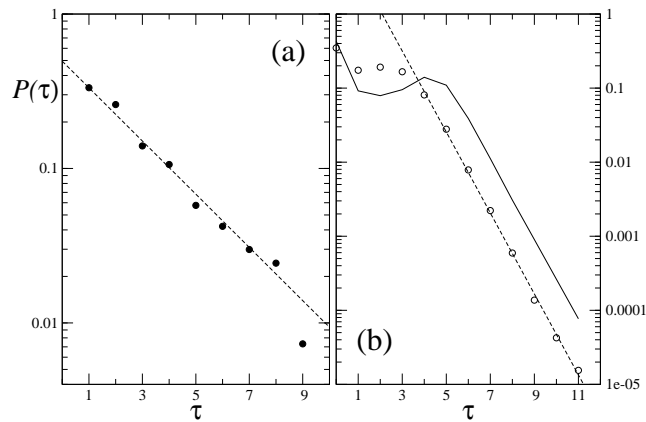}
\end{center}
\caption{\protect 
        Distribution $P(\tau)$ of spreading times $\tau$ for (a) the
        APL network of 8 generations, and (b) the real school 
        network (circles) and the BA network with $m=9$ and $N=1000$ 
        (solid line).}
\label{fig09}
\end{figure}

Another quantity of interest is the distribution $P(\tau)$ of
spreading times, which clearly decays exponentially for the Apollonian
network, as illustrated in Fig.~\ref{fig09}a.
This behavior can be also obtained analytically by considering that 
$P(\tau)d\tau=P(k)dk$ and using Eq.~(\ref{eq1}) together with the degree 
distribution, $P(k) \propto k^{-\gamma}$, to
obtain 
\begin{equation}
P(\tau) \propto e^{\tau (1-\gamma)/B} ,
\label{eq2}
\end{equation}
for large $k$. 
The slope in Fig.~\ref{fig09}a is precisely 
$(1-\gamma)/B = -0.17$ using $B$ from Fig.~\ref{fig07}a and $\gamma = 2.58$ 
from Ref.~\cite{hans}.
For the school network, $P(\tau)$ follows also an exponential decay for 
large $\tau$, but with a 3.5 times smaller characteristic decay time, 
and has a maximum for small $\tau$, as seen
in Fig.~\ref{fig09}b (circles). 
Compared to the $P(\tau)$ of the Barab\'asi-Albert network with $m=9$ 
(solid line), the shapes are similar but the Barab\'asi-Albert case
is slightly  shifted to the right, due to the larger minimal number of 
connections.

Many other regimes of gossip and of propagation phenomena can be
also addressed with these two quantities. Namely, a more realistic
scenario could be addressed by enabling each node to transfer information
with a probability $0\leq p\leq 1$. 
Further, the assumption that the person to which a gossip did not
spread at the first attempt, will never get it, yields a regime
similar to percolation conditional to the neighborhood of the victim.
Differently, if at each time-step the neighbors which already know the 
gossip repeatedly try to spread it to the common friends, one observes 
the same value of $f$ measured for $q=1$, and
the spreading time scales as $\tau^{\prime}\sim\tau/q$, where
$\tau$ is measured for $q=1$.
Finally, other possible regimes comprehend the situation where the gossip 
spreading over strangers, i.e.~over nodes which are not directly connected 
to the victim. 
Such cases are being studied in detail and results will be presented
elsewhere\cite{largefofoca}.

\section{Discussion and conclusions}
\label{sec:conclusions}

In this paper we presented and developed recent achievements in
social network research, concerning the modeling of empirical networks,
and specific mathematical tools to address their structure and 
dynamical processes on them.

Concerning the modeling of empirical networks, we described briefly a 
recent approach based on a system of mobile agents. Further developments
were given, namely in what concerns the analytical expression which fits
the typical degree distributions observed in empirical social networks.
We gave evidence that such distributions follow a Brody distribution
which depends on a single parameter that scales with the average 
degree of the network. A question which now remains to be answered is
how to derive such distribution from an analytical 
and meaningful approach.

Showing that the usual clustering coefficient is, in general, inappropriate
when addressing the clustering properties of social networks, we described
a suitable measure to access these properties and presented its additional 
applications for estimating the distribution of cycles of higher order.
This additional clustering coefficient was also put in a general framework
with different other higher-order coefficients, that could be useful
for particular situations of multipartite networks.
An expansion combining all possible coefficients was also proposed,
motivated by previous works\cite{newman03}, which depends only on the degree 
distribution and degree-degree correlations.
However, computational effort to compute such coefficients increases 
exponentially with their order and therefore it is not yet clear how useful 
such an expansion may be.

Finally, to study dynamical processes in social networks, in particular
the propagation of information, two simple measures were introduced.
Namely, a spread factor, which measures the maximal relative size of the
neighborhood reached, when the information starts from a local source
(node), and a spreading time, which gives the number of sufficient steps
to reach such maximal size.
This two measures gave rise to introduce a minimal model for gossip
propagation, which can be seen as a particular model of opinions.
Within this specific model, the spread factor was found to be minimized
by a particular non-trivial degree of the source, which is related to
the degree-degree correlations arising in the network.
If such possibility of minimizing the danger of being gossiped can be 
tested in a real situation and which other implications these findings
have in other situations - e.g.~in internet virus propagation - remain
open questions for forthcoming studies.

\section*{Acknowledgments}

The authors thank M.C.~Gonz\'alez, J.S.~Andrade Jr., L.~da Silva and 
O.~Dur\'an for useful discussions.
We thank the {\it Deutsche Forschungsgemeinschaft}, the Max Planck Price
and the {\it Volkswagenstiftung}.


\end{document}